\documentclass[prd,reprint,nofootinbib,superscriptaddress,preprintnumbers,showpacs]{revtex4-1}
\usepackage{amsmath}	
\usepackage{amssymb}	
\usepackage{amstext}	
\usepackage{latexsym}	
\usepackage{bm}	
\usepackage{graphicx,color}	
\usepackage{caption}	
\usepackage{hyperref}	
\allowdisplaybreaks[4]
\pdfoutput=1

\begin{document}
\title{The theoretical study of $D_s^+\to\pi^+\pi^0\pi^0$}
\author{Xuan Luo}
\email{xuanluo@ahu.edu.cn}
\affiliation{School of Physics and Optoelectronics Engineering, Anhui University, Hefei 230601, People's Republic of China}
\author{Aojia Xu}
\affiliation{School of Physics, Dalian University of Technology, Dalian 116024, People's Republic of China}
\author{Ruitian Li}
\affiliation{School of Physics, Dalian University of Technology, Dalian 116024, People's Republic of China}
\author{Hao Sun}
\affiliation{School of Physics, Dalian University of Technology, Dalian 116024, People's Republic of China}
\begin{abstract}
Inspired by recent BESIII’s amplitude analysis on $D_s^+\to\pi^+\pi^0\pi^0$, we develop a model based on the chiral unitary approach to discribe the $D_s^+\to\pi^+\pi^0\pi^0$ decay, where $D_s^+$ decays into a $\pi^+$ and a quark-antiquark pair through weak decay of charm quark. The interaction of pseudoscalar mesons produced by the hadronization of quark-antiquark pair will lead to dynamically generated $f_0(980)$. The dominant amplitudes from resonances $f_0(1370)$ and $f_2(1270)$ in BESIII fit are also included in our model. In addition to above contributions, to obtain a good fit we also add an amplitude from resonance $f_2(1430)$ to our model. Our model fits well $M_{\pi^0\pi^0}$ and $M_{\pi^+\pi^0}$ invariant mass distributions simultaneously. 
\end{abstract}
\maketitle
\section{introduction}
The heavy meson refers to the meson containing $b$ and $c$ quarks, whose weak decay involves the final state interaction of meson pairs and the production of scalar resonances. The study of scalar resonances begins an exploration with the mechanism of weak decay at quark level. Then the Dalitz plot and the projection of invariant mass distribution for weak decay of the heavy meson provide important information that scalar resonances are thought to be dynamically generated by the final state interaction of meson pairs. The $f_0(980)$ meson is one of these special resonances, which can be obtained by weak decay of $D_s^+$.

$f_0(980)$ meson is widely studied as a candidate for the tetraquark~\cite{Achasov:2020aun} or a quark-antiquark pair~\cite{Chen:2003za}, mainly through the processes of $D_s^+\to\pi^+\pi^+\pi^-$, $D_s^+\to\pi^+K^+K^-$ and $D_s^+\to\pi^+\pi^0\pi^0$. Ref.~\cite{Dias:2016gou} adopted the chiral unitary approach in coupled channels to study the final state interaction of two pseudoscalar mesons in $D_s^+\to\pi^+\pi^+\pi^-$ and $D_s^+\to\pi^+K^+K^-$ decays and the $f_0(980)$ signal was found in both $\pi^+\pi^-$ and $K^+K^-$ invariant mass distributions. Then Ref.~\cite{Wang:2021naf} found that, $f_0(980)$ resonance was the dominant contribution closing to the $K^+K^-$ threshold in the decay of $D_s^+\to\pi^+K^+K^-$. Later Ref.~\cite{Wang:2021ews} not only provided a full analysis of $K\bar{K}$ and $K\pi$ mass spectra but calculated the branching ratios of the dominant decay channels both in $s$-wave and $p$-wave, which had a good agreement with the experimental data from Ref.~\cite{BESIII:2020ctr}. In addition, $f_0(980)$ meson is also studied in $B$ decay, more processes about $B$ decay in~\cite{Liang:2014tia,Bayar:2014qha,Liang:2014ama,Liang:2015qva,Wang:2016wpc,Liang:2017ijf,Liang:2017lrb,Xie:2018rqv,Ahmed:2020qkv,Ahmed:2021oft}.

Other studies on $D_s^+$ decay, for the process of $D_s^+\to\pi^+\pi^0\eta$, instead of $W$-annihilation mechanism, Ref.~\cite{Molina:2019udw} proposed that the decay of $D_s^+$ occured through internal $W$-emission, where $a_0(980)$ resonance was proved to be a dynamically generated resonance by the final state interaction in coupled channels. Furthermore in Ref.~\cite{Hsiao:2019ait}, the branching ratios of $D_s^+\to\pi^+(a_0(980)\to)\pi^0\eta$ and $D_s^+\to\pi^0(a_0(980)\to)\pi^+\eta$ decays were calculated by using the triangular rescattering mechanism, which were also consistent with the experimental data. The research on $K_s^0$ also plays an important role in the discovery of resonance natures. The contribution of $a_0(1710)$ was analyzed detailedly in Ref.~\cite{Zhu:2022wzk} within the process of $D_s^+\to\pi^+K_s^0K_s^0$, which was dynamically generated by $K^*\bar{K}^*$ final state interaction and finally decayed to $K_s^0K_s^0$. The contribution of $a_0(1710)$ was also found in the process of $D_s^+\to\pi^+K^+K_s^0$~\cite{Zhu:2022duu} and finally decayed to $K^+\bar{K}^0$. 

As an isospin partner of $f_0(980)$, $a_0(980)$ has similar properties to $f_0(980)$, and there are many researches on it. For the singly Cabibbo supressed process of $D^+\to\pi^+\pi^0\eta$, the Dalitz plot was predicted in Ref.~\cite{Duan:2020vye}. Then Ref.~\cite{Ikeno:2021kzf} tested the mechanism of $W$ boson, and found that both internal and external $W$-emission mechanisms were possible, in which the final state interaction of meson pairs led to the $a_0(980)$ resonance. Similarly, Ref.~\cite{Wang:2021kka} studied the resonance signal of $a_0(980)$ within the $\pi^0\eta$ invariant mass distribution for $D^0\to\pi^0\eta\eta$ decay, and predicted the possible $f_0(500)$, $f_0(980)$, $a_0(980)$ resonances in $D^0\to\pi^0\pi^0\pi^0$ and $D^0\to\pi^0\pi^0\eta$ decays. The Cabibbo favored decay $D^0\to K^-\pi^+\eta$ was studied in Ref.~\cite{Toledo:2020zxj} by using the external $W$-emission, which obtained the $M_{K\pi}$ distribution containing the contributions of $a_0(980)$ and $K_0^*(700)$ in a good agreement with experimental data. For the doubly Cabibbo supressed process of $D^+\to K^-K^+K^+$, in Ref.~\cite{LHCb:2019tdw}, it was found that the resonant contribution mainly came from the $s$-wave component of the system. And later the resonant contribution from the $s$-wave was theoretically demonstrated in Ref.~\cite{Roca:2020lyi}, where it was explained by the mechanism of final state interaction of pseudoscalar pairs. 

Recently, BESIII Collaboration has reported a more accurate measurement of the absolute branching fraction of $D_s^+\to\pi^+\pi^0\pi^0$ decay~\cite{BESIII:2021eru} than that CLEO Collaboration has performed before~\cite{CLEO:2009vke}, and firstly performed an amplitude analysis of this process. Comparing to the decays of $D_s^+\to\pi^+\pi^+\pi^-$ and $D_s^+\to\pi^+K^+K^-$, there is no contribution of $a_0(980)\to K^+K^-$ or $\rho\to\pi^+\pi^-$ in $D_s^+\to\pi^+\pi^0\pi^0$ decay. Therefore, it can be used as a relatively clean channel to study $f_0(980)$ resonance. It is interesting that BESIII Collaboration also measured the resonance signal of $f_0(500)$ in $D_s^+\to\pi^+\pi^0\pi^0$ decay and found that the significance of $f_0(500)$ was less than $2\sigma$ while there was a pronounced peak for $f_0(980)$. Furthermore the chiral unitary approach can be used to explain this phenomenon by the final state interaction of pseudoscalar mesons, which can naturally lead to the $f_0(980)$ resonance. In the whole process, we do not introduce $f_0(500)$ or $f_0(980)$ resonance artificially, but it is dynamically generated by chiral unitary approach. This approach is an unitary extension of the chiral perturbation theory~\cite{Ananthanarayan:1996gj,Oller:1997ti,Locher:1997gr,Oller:1998hw}, characterized by the Bethe-Salpeter equation for the meson-meson interaction in coupled channel, which can provide a complete meson-meson rescattering amplitude and predict the invariant mass distribution with a minimum input.

In our work, we develop a model containing a few free parameters to fit the $M_{\pi^0\pi^0}$ and $M_{\pi^+\pi^0}$ distributions, simultaneously~\footnote{Nearly at the same time, a work \cite{Zhang:2022xpf} focus on the same process appears in the literature.}. Except the $f_0(980)$ resonance calculated by using the chiral unitary approach, we also consider the other possible contributions via the intermediate mesons in $p$-wave and $d$-wave. We wish to reproduce the peak of $f_0(980)$ resonance in $M_{\pi^0\pi^0}$ distribution without the appearance of $f_0(500)$ resonance, focus on the external $W$-emission mechanism and the final state interaction in coupled channel, so as to prove the chiral unitary approach is applicable in this process. We also provide an arguement for the necessity of amplitudes in the model that both $f_2(1430)$ and $f_2(1270)$ contributions removed from the full model have obvious influence on fit effect. The final result shows a good agreement with the experimental data, which indirectly revealing the nature of $f_0(980)$ resonance.

Our work is organized as follows. In Sec.~II, we introduce the formalism of each amplitude in our model, especially the production of $f_0(980)$ resonance within the chiral unitary approach. The combined fit to $M_{\pi^0\pi^0}$ and $M_{\pi^+\pi^0}$ distributions for $D_s^+\to\pi^+\pi^0\pi^0$ decay and discussions are presented in Sec.~III. Finally, a short conclusion is made in Sec.~IV.

\section{Formalism}
In this process, $D_s^+$ meson decays into the final state through a weak decay of charm quark, which includes three steps. First, the $c$ quark in $D_s^+$ meson converts into a $s$ quark with the $u\bar{d}$ pair produced by external $W$-emission, which is depicted in Fig.1. Note that the internal $W$-emission contributions have been ignored in this work since they are color-suppressed. We need $\pi^+\pi^0\pi^0$ in the final state, therefore, the $u\bar{d}$ pair formes the $\pi^+$ meson and the $s\bar{s}$ pair hadronizes with a $q\bar{q}(=\bar{u}u+\bar{d}d+\bar{s}s)$ pair from the vacuum. After that the pseudoscalar meson pairs produced by the hadronization will lead to dynamical generated resonances. The weak decay of $D_s^+$ can be expressed as 
\begin{equation}
	\begin{split}
		D_s^+ &\Rightarrow V_{cs}V_{ud}((u\bar{d}\to\pi^+)[s\bar{s}\to s\bar{s}\cdot(u\bar{u}+d\bar{d}+s\bar{s})])\\
		&\Rightarrow V_{cs}V_{ud}(u\bar{d}\to\pi^+)[M_{33}\to(M\cdot M)_{33}],
	\end{split}
\end{equation}
where the $M$ is the matrix consisted of the $q\bar{q}$ quarks,
\begin{equation}
	M=\left(\begin{array}{ccc}
		u\bar{u} & u\bar{d} & u\bar{s}\\
		d\bar{u} & d\bar{d} & d\bar{s}\\
		s\bar{u} & s\bar{d} & s\bar{s}
	\end{array}\right).
\end{equation}
We can also rewrite the $M$ matrix with pseudoscalar mesons,
\begin{equation}
	\Phi=M=\left(\begin{array}{ccc}
		\frac{1}{\sqrt{2}}\pi^0+\frac{1}{\sqrt{6}}\eta & \pi^+ & K^+\\
		\pi^- & -\frac{1}{\sqrt{2}}\pi^0+\frac{1}{\sqrt{6}}\eta & K^0\\
		K^- & \bar{K}^0 & -\frac{2}{\sqrt{6}}\eta
	\end{array}\right),
\end{equation}
where the $\eta'=\eta_1$ will be removed since it is unnecessary in the chiral perturbation theory, and we only take the $\eta=\eta_8$ into account. Thus the hadronization process is shown as,
\begin{equation}
	\begin{split}
		s\bar{s}(u\bar{u}+d\bar{d}+s\bar{s})&=(\Phi\cdot\Phi)_{33}\\
		&=K^-K^++K^0\bar{K}^0+\dfrac{2}{3}\eta\eta.
	\end{split}
\end{equation}
\\
\begin{figure}[htbp]
	\centering
	\includegraphics[width=0.3\textwidth]{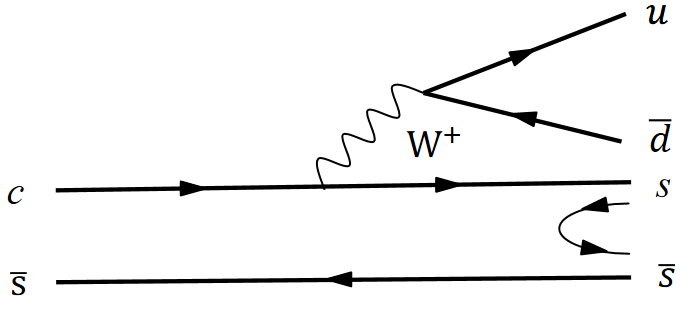}
	\captionsetup{justification=raggedright}
	\caption{The external $W$-emission for $D_s^+\to\pi^+s\bar{s}$, and $s\bar{s}$ hadronizes into the final meson pair through $q\bar{q}$ pair from the vacuum.}
\end{figure}
\begin{figure}[htbp]
	\centering
	\includegraphics[width=0.3\textwidth]{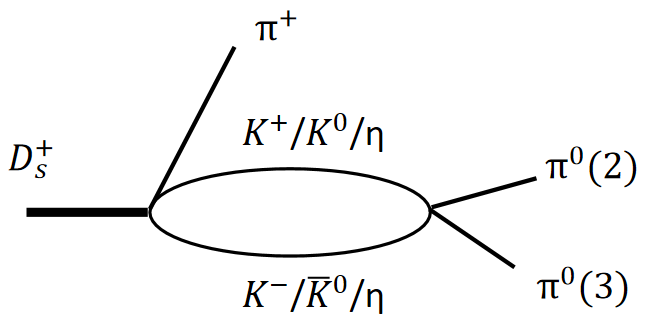}
	\captionsetup{justification=raggedright}
	\caption{Diagrammatic representation of the final states interaction, leading to the production of $\pi^0\pi^0$ final state, via the rescattering of $K^-K^+$, $K^0\bar{K}^0$ and $\eta\eta$.}
\end{figure}

Then we obtain all the final states from the hadronization of $s\bar{s}$ pair, and the $\pi^0\pi^0$ will produce through final states interaction. So we have
\begin{equation}\label{5}
	H=V_pV_{cs}V_{ud}(K^-K^++K^0\bar{K}^0+\dfrac{2}{3}\eta\eta)\pi^+,
\end{equation}
where $V_P$ is the production vertex of weak decay containing all the dynamical factors, and $V_{cs}$, $V_{ud}$ are the elements of the CKM matrix. 

In Eq.~\eqref{5}, there is no final state $\pi^0\pi^0$ produced that we want in $D_s^+$ decay. Therefore, we take the final states interaction into consideration. 
As depicted in Fig.2, one can see that the rescattering of the final states in $s$-wave, which dynamically generated the $f_0(980)$ resonance, will produce the $\pi^0\pi^0$ final state eventually. And we obtain the full amplitude of $D_s^+$ decay in $s$-wave,
\begin{equation}\label{6}
	\begin{split}	
		t(s_{23})=&V_pV_{cs}V_{ud}(G_{K^-K^+}T_{K^-K^+\to\pi^0\pi^0}\\
		&+G_{K^0\bar{K}^0}T_{K^0\bar{K}^0\to\pi^0\pi^0}+\dfrac{2}{3}\cdot2\cdot\dfrac{1}{2}G_{\eta\eta}T_{\eta\eta\to\pi^0\pi^0})\\
		=&\mathcal{D}(G_{K^-K^+}T_{K^-K^+\to\pi^0\pi^0}\\
		&+G_{K^0\bar{K}^0}T_{K^0\bar{K}^0\to\pi^0\pi^0}+\dfrac{2}{3}\cdot2\cdot\dfrac{1}{2}G_{\eta\eta}T_{\eta\eta\to\pi^0\pi^0}),
	\end{split}	
\end{equation}
where we labelled $\pi^+$, $\pi^0$, $\pi^0 $ as 1,2,3 final states and $s_{ij}=(p_i+p_j)^2$. $\mathcal{D}$ is the parameter that contains the production vertex $V_P$ and the elements of the CKM matrix $V_{cs}$, $V_{ud}$. To conform the experimental data, the normalization factor is also included in $\mathcal{D}$. In Eq.~\eqref{6}, there is a factor of 2 in the $G_{\eta\eta}T_{\eta\eta\to\pi^0\pi^0}$ term due to the two situations of the $\eta\eta\to\pi^0\pi^0$ process producing the final state. The another factor of $\frac{1}{2}$ is a result of integrating the loop function involving the indistinguishability of $\eta\eta$ mesons. As elements of the diagonal matrix, $G_{ii}$ are calculated by two intermediate meson loop functions, given by
\begin{equation}
	G_{ii}(s)=i\int\dfrac{d^4q}{(2\pi)^4}\dfrac{1}{q^2-m_1^2+i\epsilon}\dfrac{1}{(p_1+p_2-q)^2-m_2^2+i\epsilon},
\end{equation}
where $p_1$, $p_2$ are the four-momentum of the two initial particles, $q$ is one of the loop momentums of intermediate mesons, and $m_1$, $m_2$ are the masses of two intermediate mesons. Instead of using the three momentum cut-off method, we take the dimensional regularization method~\cite{Oller:2000fj} to solve this logarithmically divergent integral,
\begin{equation}
	\begin{split}
		G_{kk}(s)=&\dfrac{1}{16\pi^2}\{a_\mu+\mathrm{ln}\dfrac{m_1^2}{\mu^2}+\dfrac{m_2^2-m_1^2+s}{2s}\mathrm{ln}\dfrac{m_2^2}{m_1^2}\\
		&+\dfrac{q_{cmk}(s)}{\sqrt{s}}[\mathrm{ln}(s-(m_2^2-m_1^2)+2q_{cmk}(s)\sqrt{s})\\
		&+\mathrm{ln}(s+(m_2^2-m_1^2)+2q_{cmk}(s)\sqrt{s})\\
		&-\mathrm{ln}(-s-(m_2^2-m_1^2)+2q_{cmk}(s)\sqrt{s})\\
		&-\mathrm{ln}(-s+(m_2^2-m_1^2)+2q_{cmk}(s)\sqrt{s})]\},
	\end{split}
\end{equation}
with $q_{cmk}(s)$ is the modulus of three momentum in the center-of-mass frame corresponding to the coupled channel,
\begin{equation}
	q_{cmk}(s)=\dfrac{\lambda^{\frac{1}{2}}(s,m_1^2,m_2^2)}{2\sqrt{s}},
\end{equation}
where $\mu$ is the dimensional regularization scale, and $a_\mu$ is the subtraction constant. Following the Ref.~\cite{Oller:2000fj}, we take the values of subtraction constants in different channels, as $a_{K^+K^-}=-1.66$, $a_{K^0\bar{K}^0}=-1.66$, and $a_{\eta\eta}=-1.71$. Besides, $\lambda$ is the k$\ddot{\mathrm{a}}$ll$\acute{\mathrm{e}}$n function, represented as $\lambda(a,b,c)=a^2+b^2+c^2-2(ab+ac+bc)$.
\\
\begin{figure*}[htbp]
	\centering
	\includegraphics[width=0.8\textwidth]{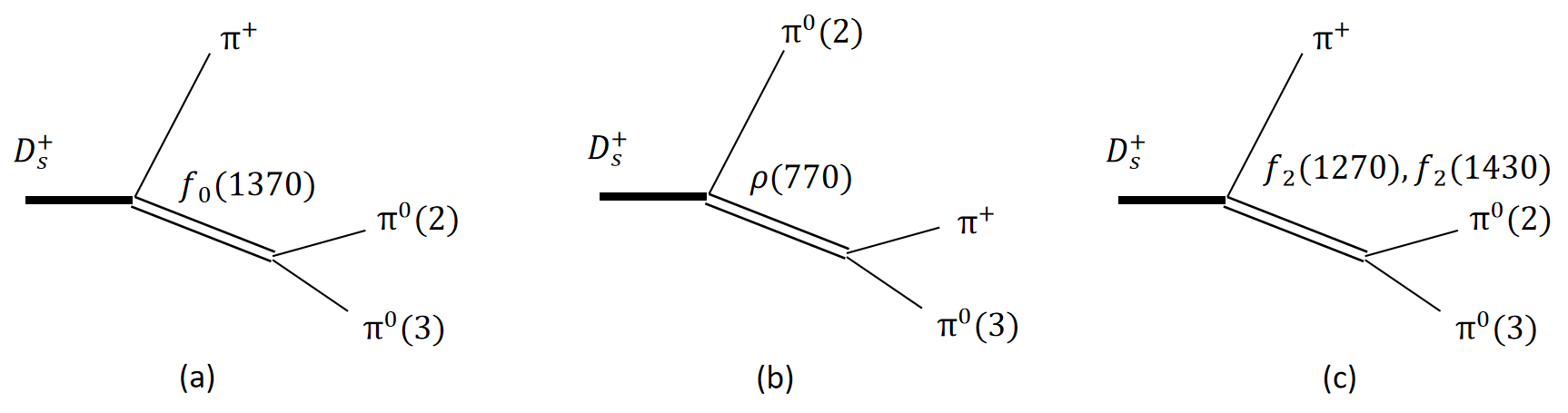}
	\captionsetup{justification=raggedright}
	\caption{The decay of $D_s^+\to\pi^+\pi^0\pi^0$ via meson resonances: (a) via $f_0(1370)$ meson in $s$-wave; (b) via $\rho(770)$ in $p$-wave; (c) via $f_2(1270)$ or $f_2(1430)$ in $d$-wave.}
\end{figure*}

In addition, the elements $t_{i\to j}$ are the scattering amplitudes making up the scattering matrix. In the chiral unitary approach, we can obtain them by calculated the coupled channel Bethe-Salpter equation~\cite{Oset:2016lyh}
\begin{equation}
	T=[1-VG]^{-1}V,
\end{equation}
where the $V$ is a $5\times5$ sysmetric matrix for scattering potential in $s$-wave with the elements are taken from Ref.~\cite{Gamermann:2006nm}, corresponding to the coupled channels. Here the coupled channels for $I=0$ are usually numbered as 1 for $\pi^+\pi^-$, 2 for $\pi^0\pi^0$, 3 for $K^+K^-$, 4 for $K^0\bar{K}^0$, and 5 for $\eta\eta$. Thus the $V$ matrix can be expressed as
\begin{equation}
	\begin{split}
		&V_{11}=-\dfrac{1}{2f^2}s, \ V_{12}=-\dfrac{1}{\sqrt{2}f^2}(s-m_\pi)^2, \ V_{13}=-\dfrac{1}{4f^2}s,\\
		&V_{14}=-\dfrac{1}{4f^2}s, \ \ V_{15}=-\dfrac{1}{3\sqrt{2}f^2}m_\pi^2, \ \ V_{22}=-\dfrac{1}{2f^2}m_\pi^2,\\
		&V_{23}=-\dfrac{1}{4\sqrt{2}f^2}s, \ \ V_{24}=-\dfrac{1}{4\sqrt{2}f^2}s, \ \ V_{25}=-\dfrac{1}{6f^2}m_\pi^2,\\
		&V_{33}=-\dfrac{1}{2f^2}s, \ \ V_{34}=-\dfrac{1}{4f^2}s,\\
		&V_{35}=-\dfrac{1}{12\sqrt{2}f^2}(9s-6m_\eta^2-2m_\pi^2), \ \ V_{44}=-\dfrac{1}{2f^2}s,\\
		&V_{45}=-\dfrac{1}{12\sqrt{2}f^2}(9s-6m_\eta^2-2m_\pi^2),\\ &V_{55}=-\dfrac{1}{18f^2}(16m_K^2-7m_\pi^2),
	\end{split}
\end{equation}
where $f=93$~MeV is the pion decay constant, $m_\pi$, $m_K$, and $m_\eta$ are the averaged masses of pion, kaon, and $\eta$ mesons, respectively. 

One thing we need to note is that, the chiral unitary approach has a limit that we can only make reliable predictions up to 1.1-1.2~GeV. Since we only interested in the region below 1.1~GeV, we can take the Eq.~(19) in Ref.~\cite{Debastiani:2016ayp} to smoothly extrapolate the $GT$ amplitude above the energy cut $\sqrt{s_{\mathrm{cut}}}=1.1$~GeV,
\begin{equation}
	G(s)T(s)=G(s_{\mathrm{cut}})T(s_{\mathrm{cut}})e^{-\alpha(\sqrt{s}-\sqrt{s_{\mathrm{cut}}})},
\end{equation}
where $\alpha$ is a parameter that we can determine it by fitting to the experimental data.

Besides, the $\pi^0\pi^0$ pair can be produnced directly in $s$-wave, and the amplitude of the process $D_s^+\to\pi^+(\pi^0\pi^0)_S$ can be simply written as
\begin{equation}
	\mathcal{M}_{(\pi^0\pi^0)_S}=\mathcal{D}_{(\pi^0\pi^0)_S}e^{i\alpha_{(\pi^0\pi^0)_S}},
\end{equation}
where $\mathcal{D}_{(\pi^0\pi^0)_S}$ and $\alpha_{(\pi^0\pi^0)_S}$ are parameters of normalization constant and phase, respectively, and will be determined by fitting to the experimental data.

In $s$-wave, we also take the possibility into account that the $s\bar{s}$ pair can form the $f_0(1370)$, and the $f_0(1370)$ will decay into the $\pi^0\pi^0$ pair. The process is depicted in the Fig.3(a) and the amplitude can be discribed by the formula as
\begin{equation}
	\mathcal{M}_{f_0(1370)}(s_{23})=\dfrac{\mathcal{D}_{f_0(1370)} e^{i\alpha_{f_0(1370)}}}{s_{12}-M_{f_0(1370)}^2+iM_{f_0(1370)}\Gamma_{f_0(1370)}},
\end{equation}
where $\mathcal{D}_{f_0(1370)}$, $\alpha_{f_0(1370)}$ are parameters of the process $D_s^+\to\pi^+f_0(1370)\to\pi^+\pi^0\pi^0$ and $\Gamma_{f_0(1370)}$ is the total width of $f_0(1370)$, taken as $\Gamma_{f_0(1370)}=300$~MeV in Ref.~\cite{ParticleDataGroup:2022pth}.

Except for the production in $s$-wave, the $\pi^0\pi^0$ pair can also produce by the decay of a vector meson in $p$-wave. We consider the amplitude for the $D_s^+\to\rho(770)^+\pi^0\to\pi^+\pi^0\pi^0$, as decipted in the Fig.3(b), which is given by~\cite{Toledo:2020zxj}
\begin{equation}
	\begin{split}
		\mathcal{M}&_{\rho}(s_{12},s_{23})\\
		=&\mathcal{D}_{\rho}e^{i\alpha_{\rho}}\left[\dfrac{s_{12}-s_{23}}{s_{13}-M_{\rho}^2+iM_{\rho}\Gamma_{\rho}}+\dfrac{s_{13}-s_{23}}{s_{12}-M_{\rho}^2+iM_{\rho}\Gamma_{\rho}}\right],
	\end{split}
\end{equation}
where $\mathcal{D}_{\rho}$ and $\alpha_{\rho}$ are parameters, and $\Gamma_{\rho}=150.2$~MeV. Although there are three $s_{ij}$ variables, only two of them are independent which fulfill the following relation
\begin{equation}
	s_{12}+s_{13}+s_{23}=M_{D_s^+}^2+m_{\pi^+}^2+m_\pi^2+m_\pi^2.
\end{equation}

Similarly, the $\pi^0\pi^0$ pair can produnce directly in $d$-wave via the process $D_s^+\to\pi^+(\pi^0\pi^0)_D$ with the amplitude written as
\begin{equation}
	\begin{split}
		&\mathcal{M}_{(\pi^0\pi^0)_D}(s_{12},s_{23})\\
		&=\mathcal{D}_{(\pi^0\pi^0)_D}e^{i\alpha_{(\pi^0\pi^0)_D}}\left[\left(\dfrac{s_{12}-2m_\pi^2}{2}\right)^2+\left(\dfrac{s_{13}-2m_\pi^2}{2}\right)^2\right],
	\end{split}
\end{equation}
where $\mathcal{D}_{(\pi^0\pi^0)_D}$ and $\alpha_{(\pi^0\pi^0)_D}$ are parameters.

In $d$-wave, the $\pi^0\pi^0$ production will proceed via the decay of $f_2(1270)$ and $f_2(1430)$, as shown in Fig.3(c). The decay amplitude is
\begin{equation}
	\begin{split}
		\mathcal{M}&_{f_2}(s_{12},s_{23})\\
		=&\dfrac{\mathcal{D}_{f_2}e^{i\alpha_{f_2}}}{s_{23}-M_{f_2}^2+iM_{f_2}\Gamma_{f_2}}\dfrac{1}{12M_{f_2}^2}\{M_{f_2}^2[-16m_\pi^4+4m_\pi^2s_{23}\\
		&+3(s_{12}-s_{13})^2]+(4m_\pi^2-s_{23})(-4m_\pi^2+s_{12}+s_{13})^2\},
	\end{split}
\end{equation}
where $\mathcal{D}_{f_2}$ and $\alpha_{f_2}$ are the parameters of the decay of $f_2(1270)$ or $f_2(1430)$. And $\Gamma_{f_2}$ is the total width, taken as $\Gamma_{f_2(1270)}=1275.5$~MeV and $\Gamma_{f_2(1430)}=1430$~MeV.

With the consideration above, we obtain the total amplitude of $D_s^+\to\pi^+\pi^0\pi^0$ as
\begin{equation}\label{19}
	\begin{split}
		t'(s_{12},s_{23})=&t(s_{23})+\mathcal{M}_{(\pi^0\pi^0)_S}(s_{12},s_{23})+\mathcal{M}_{f_0(1370)}(s_{23})\\
		&+\mathcal{M}_{\rho(770)^+}(s_{12},s_{23})+\mathcal{M}_{(\pi^0\pi^0)_D}(s_{12},s_{23})\\
		&+\mathcal{M}_{f_2(1270)}(s_{12},s_{23})+\mathcal{M}_{f_2(1430)}(s_{12},s_{23}).
	\end{split}
\end{equation}

Finally, considering that there are only two independent invariant masses, we use the formula of double differential width for three body decay~\cite{ParticleDataGroup:2022pth} as
\begin{equation}
	\dfrac{d^2\Gamma}{dM_{12}dM_{23}}=\dfrac{1}{(2\pi)^3}\dfrac{1}{2}\dfrac{M_{12}M_{23}}{8m_{D_s^+}^3}|t'(s_{12},s_{23})|^2.
\end{equation}
Then one can get the single differential mass distribution by integrating over another invariant mass. To integrate over $M_{23}$, the limits of integration in PDG~\cite{ParticleDataGroup:2022pth} are given by
\begin{equation}
	(M_{23})_{\mathrm{max}}^2=(E_2^*+E_3^*)^2-(\sqrt{E_2^{*2}-m_2^2}-\sqrt{E_3^{*2}-m_3^2})^2,
\end{equation}
\begin{equation}
	(M_{23})_{\mathrm{min}}^2=(E_2^*+E_3^*)^2-(\sqrt{E_2^{*2}-m_2^2}+\sqrt{E_3^{*2}-m_3^2})^2,
\end{equation}
where the $E_2^*$ and $E_3^*$ are energies in the $\pi^+\pi^0$ rest frame,
\begin{equation}
	E_2^*=\dfrac{s_{12}-m_1^2+m_2^2}{2\sqrt{s_{12}}},
\end{equation}
\begin{equation}
	E_3^*=\dfrac{m_{D_s^+}^2-s_{12}-m_3^2}{2\sqrt{s_{12}}}.
\end{equation}
And to integrate over $M_{12}$, the limits of integration are given by
\begin{equation}
	(M_{12})_{\mathrm{max}}^2=(E_2^*+E_1^*)^2-(\sqrt{E_2^{*2}-m_2^2}-\sqrt{E_1^{*2}-m_1^2})^2,
\end{equation}
\begin{equation}
	(M_{12})_{\mathrm{min}}^2=(E_2^*+E_1^*)^2-(\sqrt{E_2^{*2}-m_2^2}+\sqrt{E_1^{*2}-m_1^2})^2,
\end{equation}
where the $E_2^*$ and $E_1^*$ are energies in the $\pi^0\pi^0$ rest frame,
\begin{equation}
	E_2^*=\dfrac{s_{23}-m_3^2+m_2^2}{2\sqrt{s_{23}}},
\end{equation}
\begin{equation}
	E_1^*=\dfrac{m_{D_s^+}^2-s_{23}-m_1^2}{2\sqrt{s_{23}}}.
\end{equation}

\section{results}
According to our calculation above, we have introduced seven amplitudes with a total of 14 parameters. In Fig.4, we determine the parameters by simultaneously fitting the $M_{\pi^0\pi^0}$ and $M_{\pi^+\pi^0}$ distributions of $D_s^+\to\pi^+\pi^0\pi^0$ from the BESIII data using the model above, and the result is shown in Table~I. Different from the experiment, two additional contributions, $\rho(770)^+$ and $f_2(1430)$, have been introduced in our model. While the branching ratios of them are smaller than that of the main contributions and are ignored in the experimental figures. For a more meaning comparsion with experimental data, we need to smear the theoretical curve with experimental resolution or bin width. We have smeared the theoretical curve with bin width and choosed the Gaussian function for smearing which is a good choice. For a original function $f(x)$, the smeared function $\tilde{f}(x)$ is
\begin{eqnarray}
	\begin{aligned}
		\tilde{f}(x)=\int\limits_{y_{\rm min}}^{y_{\rm max}}dy \ g(|x-y|)f(y),
	\end{aligned}
\end{eqnarray} 
with
\begin{eqnarray}
	\begin{aligned}
		g(x)=\frac{1}{\sqrt{\pi}\sigma}\exp\left( -\frac{x^2}{\sigma^2} \right),
	\end{aligned}
\end{eqnarray} 
where $\sigma=\frac{\sqrt{2}}{2}\Delta$, and $\Delta$ is bin width.
The ranges of integration are (after checking covergence)
\begin{eqnarray}
	\begin{aligned}
		&y_{\rm min} \sim x-3\sigma,\\
		&y_{\rm max} \sim x+3\sigma.
	\end{aligned}
\end{eqnarray}
\begin{table}[htbp]
	\renewcommand{\arraystretch}{1.8}
	\tabcolsep=3.mm
	\captionsetup{justification=raggedright}
	\caption*{TABLE~I. Parameters and fit values for the full model obtained from fitting the BESIII data~\cite{BESIII:2021eru}.}
	\begin{tabular}[b]{cccc}
		Parameters & Fit values & Parameters & Fit values \\
		$\mathcal{D}$ & 783.1 & $\alpha$ & -0.974~$\mathrm{GeV^{-1}}$ \\
		$\mathcal{D}_{f_0(1370)}$ & 253.4~$\mathrm{GeV^2}$ & $\alpha_{f_0(1370)}$ & 3.535 \\
		$\mathcal{D}_{(\pi^0\pi^0)_S}$ & 306.5 & $\alpha_{(\pi^0\pi^0)_S}$ & -4.700 \\
		$\mathcal{D}_{\rho}$ & -12.15 & $\alpha_{\rho}$ & 9.507 \\
		$\mathcal{D}_{f_2(1270)}$ & 150.0~$\mathrm{GeV^{-2}}$ & $\alpha_{f_2(1270)}$ & -8.803 \\
		$\mathcal{D}_{f_2(1430)}$ & 166.7~$\mathrm{GeV^{-2}}$ & $\alpha_{f_2(1430)}$ & 4.368 \\
		$\mathcal{D}_{(\pi^0\pi^0)_D}$ & -29.75~$\mathrm{GeV^{-4}}$ & $\alpha_{(\pi^0\pi^0)_D}$ & 1.002
	\end{tabular}
\end{table}

For the region above 1.1~GeV in $s$-wave, we smoothly extrapolate the amplitude above the energy cut instead of cut-off directly. The impact of the model on the uncertainty below 1.1~GeV in $s$-wave can be referred to the discussions in Ref.~\cite{Guo:2005wp,Debastiani:2016ayp}, and the influence on $f_0(980)$ below 1.1~GeV is very small. Therefore, we can take the smoothing extrapolation parameter as another parameter and obtain it by fitting to the experimental data.
\begin{figure}[htbp]
	\centering
	\includegraphics[width=0.48\textwidth]{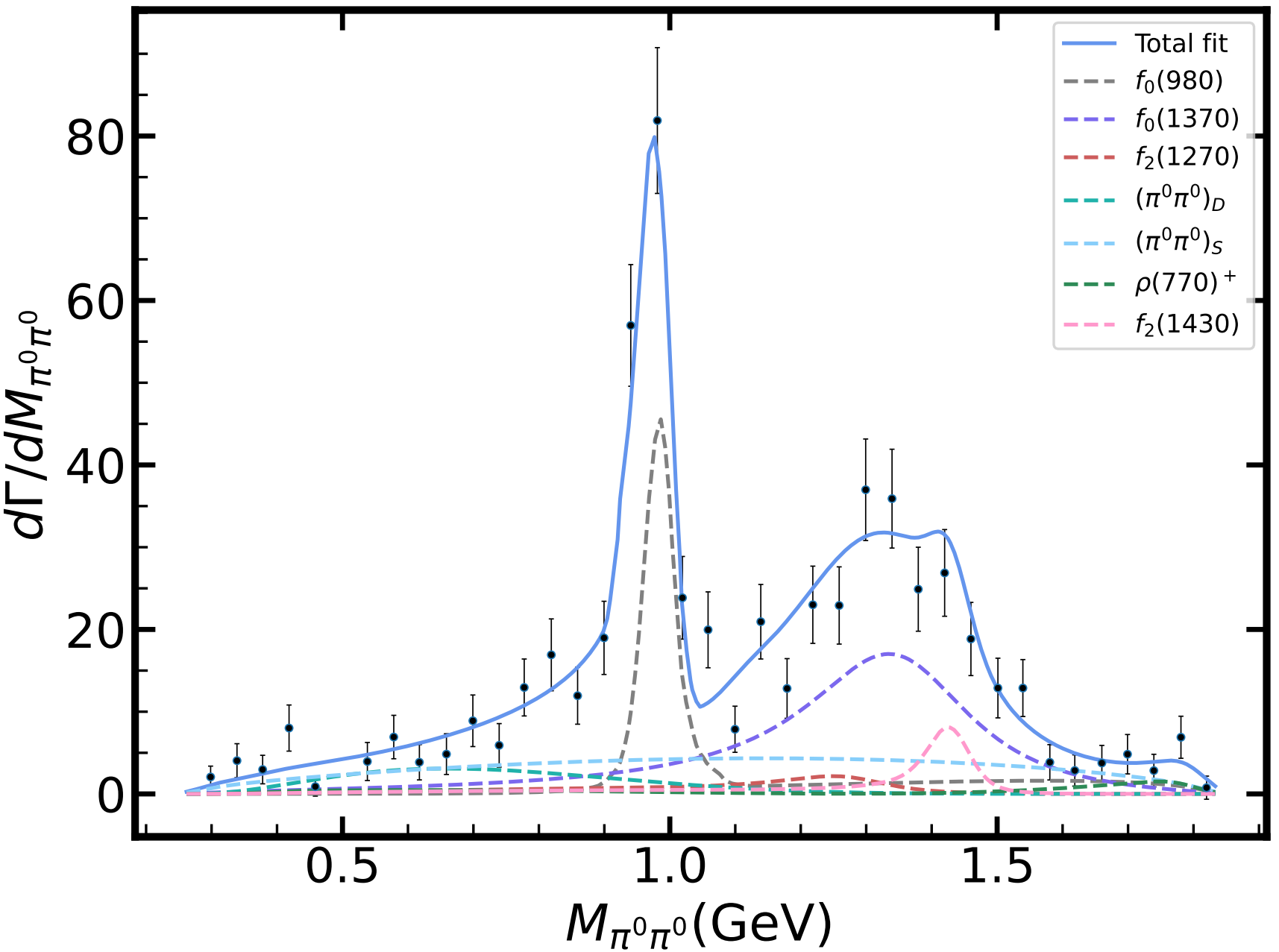}
	\captionsetup{justification=raggedright}
	\caption*{(a) $M_{\pi^0\pi^0}$ distribution for $D_s^+\to\pi^+\pi^0\pi^0$ decay with $\chi^2/dof.=54.63/38=1.44$.}\label{Fig.4(a)}
	\centering
	\includegraphics[width=0.48\textwidth]{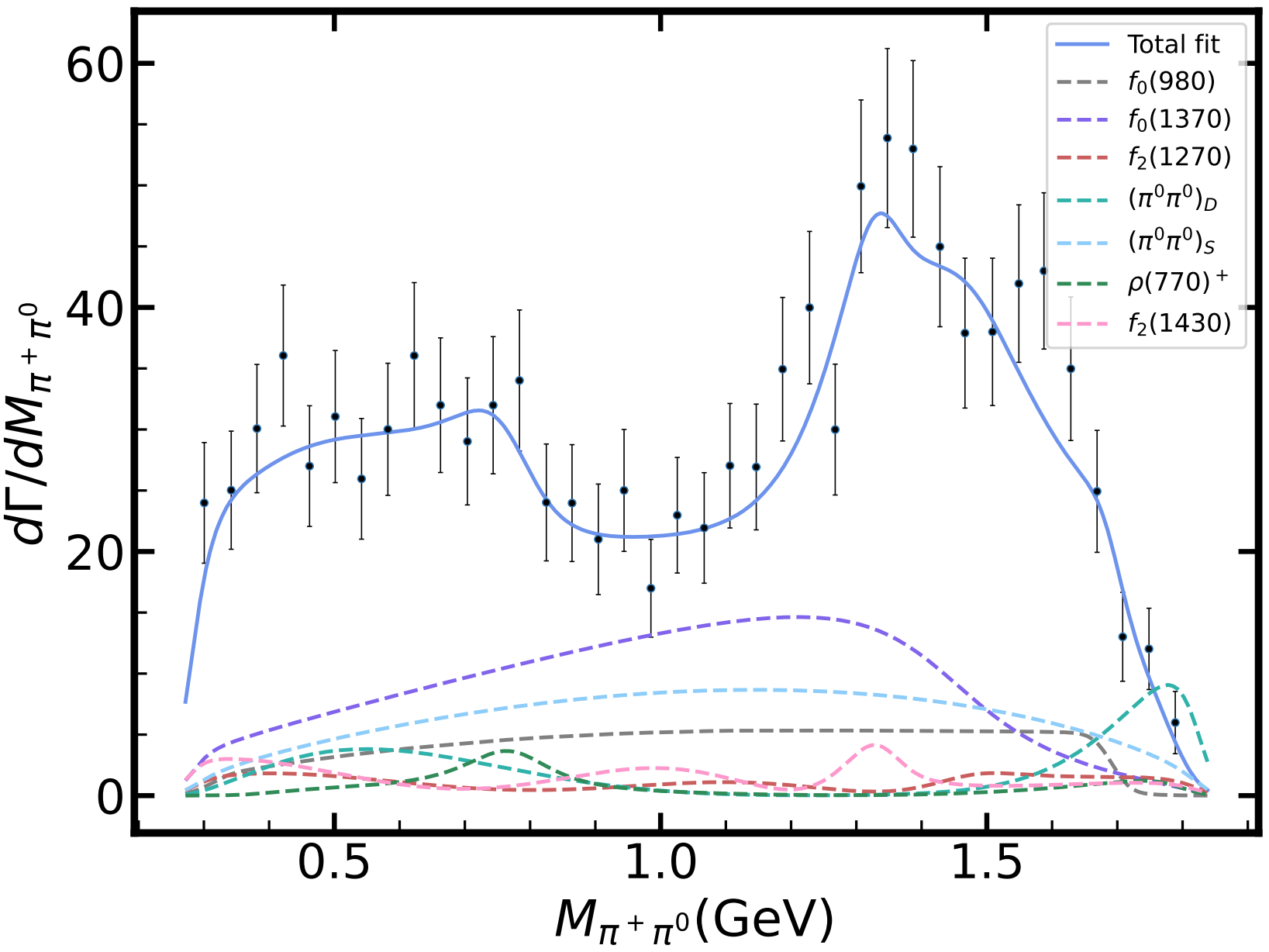}
	\captionsetup{justification=raggedright}
	\caption*{(b) $M_{\pi^+\pi^0}$ distribution for $D_s^+\to\pi^+\pi^0\pi^0$ decay with $\chi^2/dof.=27.33/38=0.72$.}\label{Fig.4(b)}
	\caption{Combined fit to $M_{\pi^0\pi^0}$ and $M_{\pi^+\pi^0}$ distributions for $D_s^+\to\pi^+\pi^0\pi^0$ decay. The curve labeled as 'Total fit' shows the result of full model. The dashed curves are the results of component contributions. The data are from Ref.~\cite{BESIII:2021eru}.}\label{Fig.4}
\end{figure}

In Fig.4(a), we present the $M_{\pi^0\pi^0}$ distribution of $D_s^+$ decay. One can clearly see that the peak at 0.98~GeV, corresponding to the dynamically generated resonance $f_0(980)$. On the basis of the original model, we take into account the experimental resolution in energy, so that the value of the $f_0(980)$ peak is in better agreement with the experiment. Another noteworthy point is that the contribution of $f_0(1370)$ dominates at $M_{\pi^0\pi^0}$ higher than 1.1~GeV, while the contribution of $f_2(1270)$ is not sharp. Except for a discrepancy around 1.43~GeV, the overall shape of invariant mass distribution is in good agreement with experimental data, since the additional $f_2(1430)$ contribution leads to a small peak at 1.43~GeV in $M_{\pi^0\pi^0}$ distribution. Then in Fig.4(b), we depict the $M_{\pi^+\pi^0}$ distribution. Obviously $f_0(1370)$ is the main contribution, and $f_2(1430)$ has a certain impact on the peak of the total fit curve around 1.35~GeV.
\begin{figure}[htbp]
	\centering
	\includegraphics[width=0.48\textwidth]{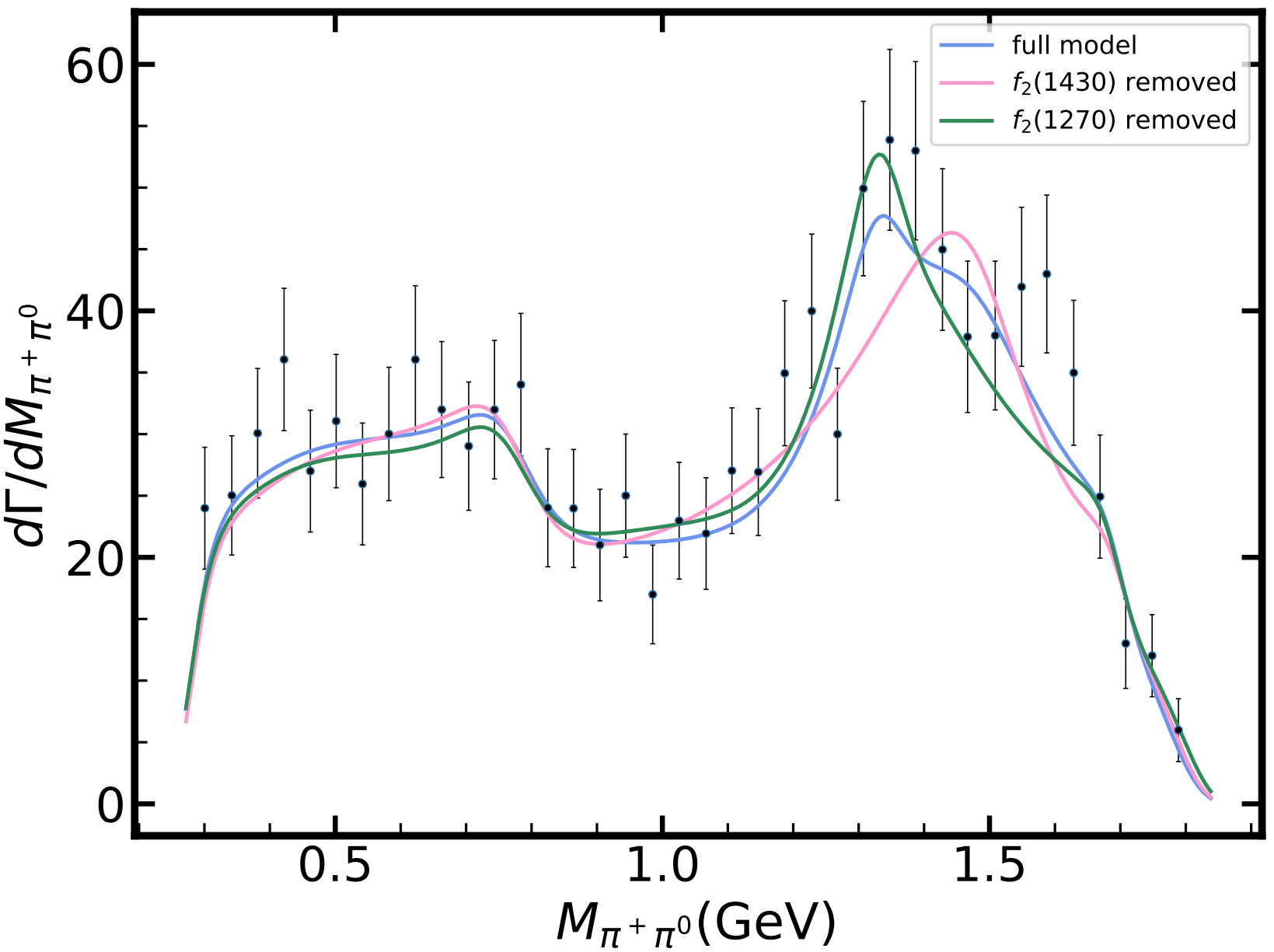}
	\captionsetup{justification=raggedright}
	\caption{$M_{\pi^+\pi^0}$ distribution from combined fits for different models with blue curve for the full model, pink curve for the full model with $f_2(1430)$ contribution removed and green curve for the full model with $f_2(1270)$ contribution removed. The data are from Ref.~\cite{BESIII:2021eru}.}
\end{figure}

In Fig.5, we present the $M_{\pi^+\pi^0}$ distribution for the full model with $f_2(1430)$ or $f_2(1270)$ contribution removed. In the region of 1.3-1.6~GeV, the curve is significantly lower than the experimental data point if the $f_2(1430)$ contribution is removed from the full model, and the fit effect of curve near the edge of phase space is worse than that with $f_2(1430)$ contribution added. Therefore, the $f_2(1430)$ contribution is necessary in our model. For the contribution of $f_2(1270)$, the amplitude is very small both in Fig.4(a) and Fig.4(b). However if we remove the $f_2(1270)$ contribution, one can see that, although the total fit curve is very close to the experimental data at about 1.35~GeV, it is not ideal in the region of 1.4-1.7~GeV. Thus, in our model, we chose to add the $f_2(1270)$ contribution to make the fit effect more consistent with the experimental data. Other contributions appeared in the experiment, $(\pi^+\pi^0)_D$, $\rho(1450)$ and $f_0(1500)$, which we have tested but the impact on fit effect is so slight that we do not add them into our model.

\begin{figure}[htbp]
	\centering
	\includegraphics[width=0.48\textwidth]{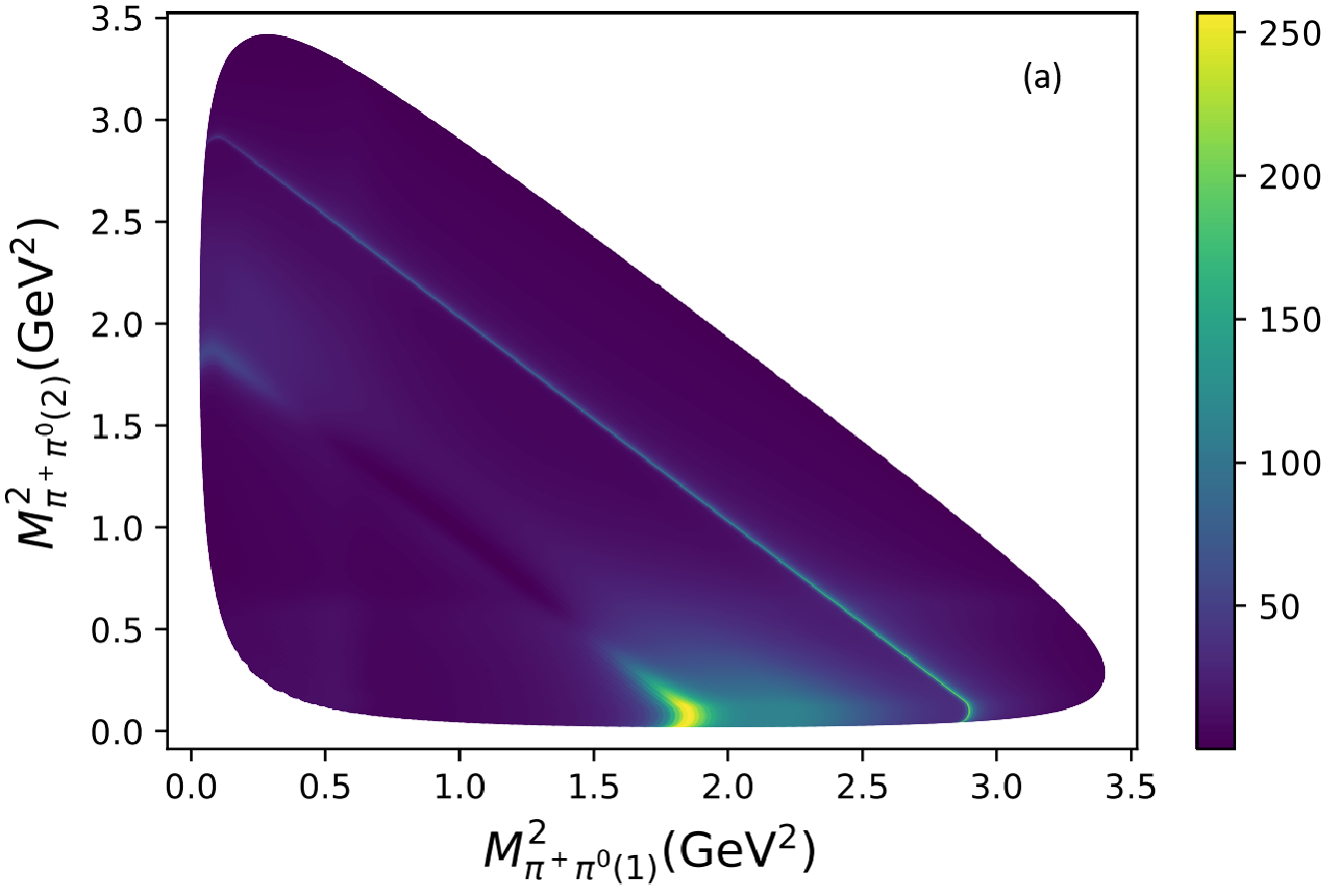}
	\includegraphics[width=0.48\textwidth]{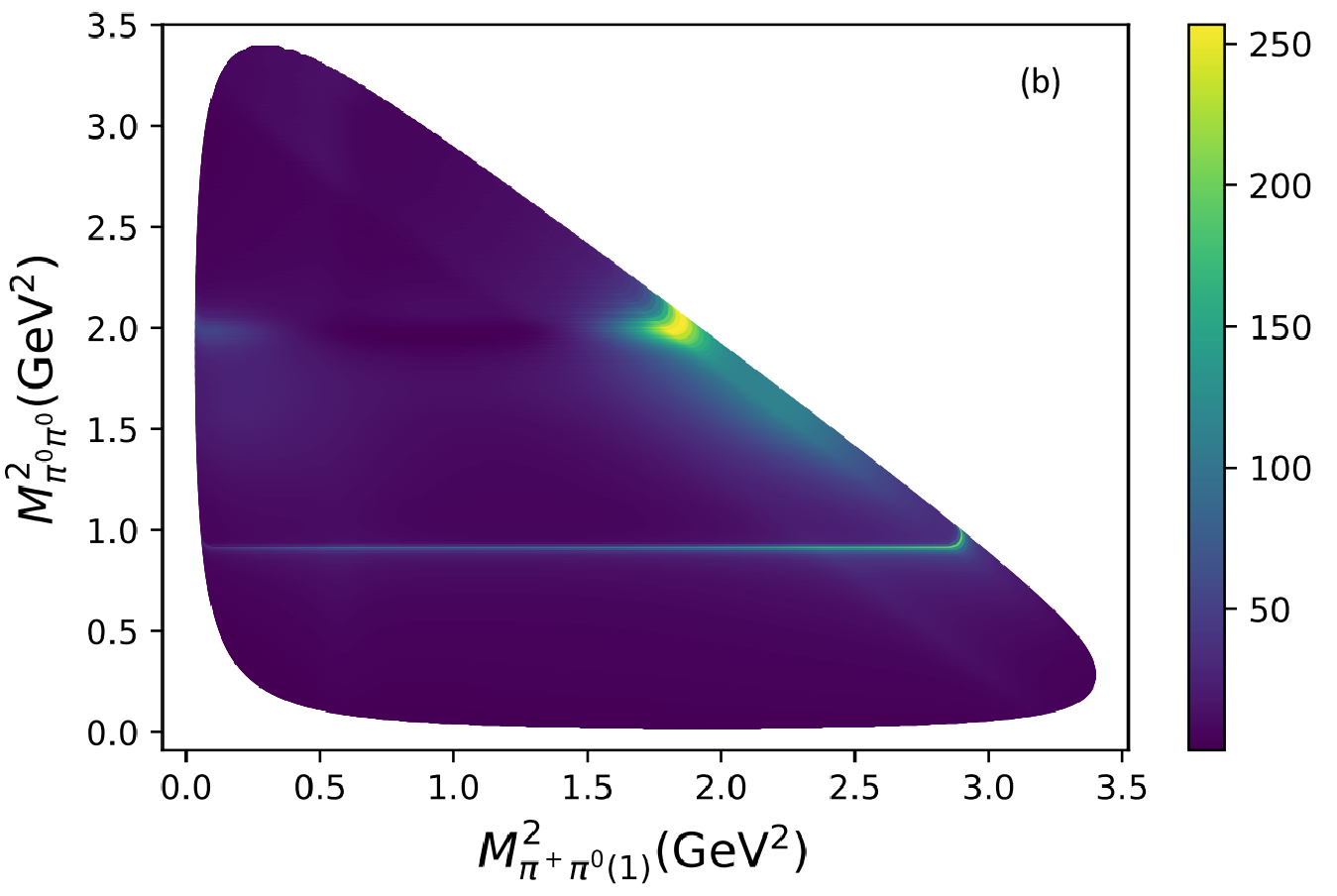}
	\caption{The Dalitz plots for the $D_s^+\to\pi^+\pi^0\pi^0$ decay.}
\end{figure}
We also show the Dalitz plots of "$M_{\pi^0\pi^0}^2$" versus "$M_{\pi^+\pi^0(1)}^2$" and "$M_{\pi^+\pi^0(2)}^2$" versus "$M_{\pi^+\pi^0(1)}^2$" in the $D_s^+$ process, depicted in Fig.6, which can be used to test the model.

\section{conclusion}
We have made a theoretical study of $D_s^+\to\pi^+\pi^0\pi^0$. In our model, we consider the external $W$-emission at quark level and hadronize with $q\bar{q}$ pairs to produce pseudoscalar mesons, which finally decay into $\pi^+\pi^0\pi^0$ state. According to the chiral unitary approach, we focus on the interaction of pseudoscalar mesons, where $f_0(980)$ can be explain as a dynamically generated state in $s$-wave. 
We also take contributions from $p$-wave and $d$-wave resonances into account.
In order to explore the consistency of our approach and experiment measurements, we investigate the $M_{\pi^0\pi^0}$ and $M_{\pi^+\pi^0}$ invariant mass distribution, using the parameters mentioned in formalism to fit the experimental data, and found a clear peak of $f_0(980)$ in $M_{\pi^0\pi^0}$ invariant mass distribution with no contribution of $f_0(500)$ appeared, which conform to the theoretical prediction. Therefore, the chiral unitary approach is well applicable this kind of processes, providing support for explaining the nature of $f_0(980)$. Furthermore, we compared the full model with the model with $f_2(1430)$ or $f_2(1270)$ contribution removed to verify the necessity of amplitudes in our model and find that only the full model fit the experimental data best. 

\section{acknowledgments}
XL is supported by the National Natural Science Foundation of China under Grant No. 12205002. HS is supported by the National Natural Science Foundation of China (Grant No.12075043, No. 12147205).

\bibliographystyle{unsrt}
\bibliography{a}
\end{document}